\documentstyle[12pt]{article}
\newlength{\extraspace}
\newlength{\extraspaces}
\catcode`\@=11
\def\numberbysection{\@addtoreset{equation}{section}
\def\theequation{\arabic{section}.\arabic{equation}}}
\newcommand{\be}{\begin{equation}
\addtolength{\abovedisplayskip}{\extraspaces}
\addtolength{\belowdisplayskip}{\extraspaces}
\addtolength{\abovedisplayshortskip}{\extraspace}
\addtolength{\belowdisplayshortskip}{\extraspace}}
\newcommand{\ee}{\end{equation}}
\newcommand{\ba}{\begin{eqnarray}
\addtolength{\abovedisplayskip}{\extraspaces}
\addtolength{\belowdisplayskip}{\extraspaces}
\addtolength{\abovedisplayshortskip}{\extraspace}
\addtolength{\belowdisplayshortskip}{\extraspace}}
\newcommand{\ea}{\end{eqnarray}}
\newcommand{\newsection}[1]{
\vspace{7mm}
\pagebreak[3]
\addtocounter{section}{1}
\setcounter{subsection}{0}
\setcounter{footnote}{0}
\begin{center}
{\large {\bf \thesection. #1}}
\end{center}
\nopagebreak
\medskip
\nopagebreak
\hspace{3mm}}
\newcommand{\nonu}{\nonumber \\[.5mm]}
\newcommand{\A}{&\!\!\!}

\newcommand{\VEV}[1]{\left\langle {#1} \right\rangle}

\newcommand{\slA}{A \!\!\! \raisebox{0.2ex}{/}}

\begin{document}
\addtolength{\baselineskip}{.7mm}
\thispagestyle{empty}
\begin{flushright}
OCHA--PP--127 \\
STUPP--98--155 \\
{\tt hep-th/9810148} \\ 
October, 1998
\end{flushright}
\vspace{7mm}
\begin{center}
{\Large{\bf Supersymmetry in the AdS/CFT \\[2mm]
Correspondence 
}} \\[20mm] 
{\sc Madoka Nishimura}%\footnote{\tt e-mail: } 
\\[3mm]
{\it Department of Physics, Ochanomizu University \\
2-1-1, Otsuka, Bunkyo-ku, Tokyo 112-0012, Japan} \\[3mm]
and \\[3mm]
{\sc Yoshiaki Tanii}\footnote{
\tt e-mail: tanii@th.phy.saitama-u.ac.jp} \\[3mm]
{\it Physics Department, Faculty of Science \\
Saitama University, Urawa, Saitama 338-8570, Japan} \\[20mm]
{\bf Abstract}\\[7mm]
{\parbox{13cm}{\hspace{5mm}
We study how local symmetry transformations of ($p, q$) anti de 
Sitter supergravities in three dimensions act on fields on the 
two-dimensional boundary. The boundary transformation laws are shown 
to be the same as those of two-dimensional ($p, q$) conformal 
supergravities for $p, q \leq 2$. Weyl and super Weyl transformations 
are generated from three-dimensional general coordinate and super 
transformations. 
}}
\end{center}
\vfill
\newpage
\setcounter{section}{0}
\setcounter{equation}{0}
%\numberbysection
%
%%%%%%%  Section 1  %%%%%%%%%%%%%%%%%%%%%%%%%%%%%%%%%%%%%%%%
%
\newsection{Introduction}
It was conjectured in ref.\ \cite{MAL} that the string/M-theory 
in ($d+1$)-dimensional anti de Sitter (AdS) space times a compact 
space is equivalent to a $d$-dimensional conformal field theory 
(CFT). More precise form of this AdS/CFT correspondence was given 
in refs.\ \cite{GKP}, \cite{WITTEN}. 
According to refs.\ \cite{GKP}, \cite{WITTEN} the CFTs are defined 
on the boundary of the AdS space and 
the generating functional of operators ${\cal O}(x)$ in the 
boundary CFTs is given by the partition function of the 
string/M-theory. When the string/M-theory is represented by 
a low energy effective supergravity and the partition function 
is approximated by a stationary point of the supergravity action 
$S_{\rm SUGRA}$, one obtains a relation 
\be
\VEV{\exp \left(i \int d^d x \, \phi_0(x) 
{\cal O}(x) \right)}_{\rm CFT} 
= \exp(iS_{\rm SUGRA}[\phi]). 
\ee
Here, $\phi_0$ on the left hand side are arbitrary functions 
defined on the $d$-dimensional boundary while $\phi$ on the right 
hand side are the solutions of field equations in the bulk 
satisfying boundary conditions $\phi = \phi_0$. 
For fields satisfying the first order field equations such as 
a spinor field one should impose boundary conditions on only half 
of the components of the fields \cite{HS}, \cite{AF}. 
\par
The purpose of this paper is to study how local symmetry 
transformations in the bulk theories act on the fields $\phi_0$ 
on the boundary. We are especially interested in how local 
supertransformations act on $\phi_0$. The fields $\phi_0$ on 
the boundary are expected to form multiplets of $d$-dimensional 
conformal supergravities \cite{FFZ}, \cite{LT}. 
We consider three-dimensional ($p, q$) AdS supergravities of 
Ach\'ucarro and Townsend \cite{AT} in the bulk as simple examples. 
The AdS/CFT correspondence for three-dimensional AdS space 
was previously discussed from other points of view in 
refs.\ \cite{BH}, \cite{BBCHO}, \cite{BOER}, \cite{GKS}. 
\par
We partially fix the gauge for the local symmetries in the bulk 
and obtain residual symmetries, which preserve the gauge fixing 
conditions. 
These residual symmetry transformations act on the fields 
non-locally in the bulk. However, they can act on the boundary 
fields $\phi_0$ locally. It is shown that the transformations of 
the boundary fields have a local form for $p, q \leq 2$. 
In particular, the supertransformations in the bulk become 
two-dimensional super and super Weyl transformations on the boundary, 
while general coordinate transformations in the bulk 
become general coordinate and Weyl transformations on the boundary. 
These transformation laws are shown to be exactly the same as those of 
two-dimensional ($p, q$) conformal supergravities, i.e., conformal 
supergravities with $p$ supersymmetries of positive chirality and 
$q$ of negative chirality. 
\par
%
%%%%%%%  Section 2  %%%%%%%%%%%%%%%%%%%%%%%%%%%%%%%%%%%%%%%%
%
\newsection{Three-dimensional AdS supergravities}
The field content of the three-dimensional ($p, q$) AdS 
supergravity \cite{AT} is a dreibein $e_M{}^A$, 
Majorana Rarita-Schwinger fields $\psi_M^i$, $\psi_M^{i'}$ and 
SO($p$) $\times$ SO($q$) Chern-Simons gauge fields 
$A_M^{ij} = - A_M^{ji}$, $A_M^{i'j'} = - A_M^{j'i'}$, 
where $i, j, \cdots = 1, \cdots, p$; $i', j', \cdots = 1, \cdots, q$. 
We denote three-dimensional world indices as 
$M, N, \cdots = 0, 1, 2$ and local Lorentz 
indices as $A, B, \cdots = 0, 1, 2$. Our conventions are as follows. 
The flat metric is $\eta_{AB} = {\rm diag}(-1, +1, +1)$ and 
the totally antisymmetric tensor $\epsilon^{ABC}$ is chosen as 
$\epsilon^{012} = +1$. $2 \times 2$ gamma matrices $\gamma_A$ 
satisfy $\{\gamma_A, \gamma_B \} = 2 \eta_{AB}$. 
$\gamma$'s with multiple indices are antisymmetrized products of 
gamma matrices with unit strength. In particular, we have 
$\gamma^{ABC} = - \epsilon^{ABC}$ in three dimensions. 
The Dirac conjugate of a spinor $\psi$ 
is defined as $\bar\psi = \psi^\dagger i \gamma^0$. 
All components of gamma matrices are chosen to be real and 
Majorana spinors have two real components. 
\par
The Lagrangian is given by 
\ba
{\cal L}
\A = \A {1 \over 8\pi G} \biggl[ 
{1 \over 2} e R + 4 m^2 e 
+ {1 \over 2} i \epsilon^{MNP} \bar\psi_M^i {\cal D}_N \psi_P^i 
+ {1 \over 2} i m e \bar\psi_M^i \gamma^{MN} \psi_N^i \nonu
\A \A + {1 \over 2} i \epsilon^{MNP} \bar\psi_M^{i'} 
{\cal D}_N \psi_P^{i'} - {1 \over 2} i m e \bar\psi_M^{i'} 
\gamma^{MN} \psi_N^{i'} \nonu
\A \A - {1 \over 4m} \epsilon^{MNP} \left( 
A_M^{ij} \partial_N A_P^{ji} 
+ {2 \over 3} A_M^{ij} A_N^{jk} A_P^{ki} \right) \nonu
\A \A + {1 \over 4m} \epsilon^{MNP} \left( 
A_M^{i'j'} \partial_N A_P^{j'i'} 
+ {2 \over 3} A_M^{i'j'} A_N^{j'k'} A_P^{k'i'} \right) 
\biggr], 
\label{ATaction}
\ea
where $m$ is a positive constant. The cosmological constant is 
proportional to $m^2$. In the following we will put the 
gravitational constant as $8\pi G = 1$. 
Our conventions for the curvature tensors are 
\ba
R \A = \A e_A{}^M e_B{}^N R_{MN}{}^{AB}, \nonu
R_{MN}{}^{AB} \A = \A \partial_M \omega_N{}^{AB} 
+ \omega_M{}^A{}_C \omega_N{}^{CB} - (M \leftrightarrow N) 
\ea
and the covariant derivatives are defined as 
\ba
{\cal D}_M \psi^i_N 
\A = \A \left( \partial_M + {1 \over 4} \omega_M{}^{AB} 
\gamma_{AB} \right) \psi_N^i 
+ A_M^{ij} \psi_N^j, \nonu
{\cal D}_M \psi^{i'}_N 
\A = \A \left( \partial_M + {1 \over 4} \omega_M{}^{AB} 
\gamma_{AB} \right) \psi_N^{i'} 
+ A_M^{i'j'} \psi_N^{j'}. 
\ea
The covariant derivatives without SO($p$) $\times$ SO($q$) connection 
terms are denoted as $D_M$. The spin connection is given by 
\ba
\omega_M{}^{AB} 
\A = \A \omega_M{}^{AB}(e) 
+ {1 \over 4} i ( \bar\psi_M^i \gamma_A \psi_B^i 
- \bar\psi_M^i \gamma_B \psi_A^i 
+ \bar\psi_A^i \gamma_M \psi_B^i \nonu
\A \A + \bar\psi_M^{i'} \gamma_A \psi_B^{i'} 
- \bar\psi_M^{i'} \gamma_B \psi_A^{i'} 
+ \bar\psi_A^{i'} \gamma_M \psi_B^{i'} ), 
\label{sconnection}
\ea
where $\omega_M{}^{AB}(e)$ is the spin connection without torsion. 
%The spin connection (\ref{sconnection}) has a torsion 
%\be
%D_M e_N{}^A - D_N e_M{}^A 
%= {1 \over 2} i ( \bar\psi_M^i \gamma^A \psi_N^i 
%+ \bar\psi_M^{i'} \gamma^A \psi_N^{i'} ). 
%\ee
If $\omega_M{}^{AB}$ is treated as an independent variable in the 
Lagrangian (\ref{ATaction}), its field equation is solved by 
eq.\ (\ref{sconnection}). 
\par
The Lagrangian (\ref{ATaction}) is invariant under the following 
local transformations for arbitrary ($p, q$) up to total 
derivative terms: 
\ba
\delta e_M{}^A 
\A = \A \xi^N \partial_N e_M{}^A + \partial_M \xi^N e_N{}^A 
- \lambda^A{}_B e_M{}^B 
+ {1 \over 2} i \left( \bar\epsilon^i \gamma^A \psi_M^i 
+ \bar\epsilon^{i'} \gamma^A \psi_M^{i'} \right), \nonu
\delta \psi_M^i 
\A = \A \xi^N \partial_N \psi_M^i + \partial_M \xi^N \psi_N^i 
- {1 \over 4} \lambda^{AB} \gamma_{AB} \psi_M^i 
- \theta^{ij} \psi_M^j 
+ {\cal D}_M \epsilon^i + m \gamma_M \epsilon^i, \nonu
\delta \psi_M^{i'} 
\A = \A \xi^N \partial_N \psi_M^{i'} + \partial_M \xi^N \psi_N^{i'} 
- {1 \over 4} \lambda^{AB} \gamma_{AB} \psi_M^{i'} 
- \theta^{i'j'} \psi_M^{j'} 
+ {\cal D}_M \epsilon^{i'} - m \gamma_M \epsilon^{i'}, \nonu
\delta A_M^{ij} 
\A = \A \xi^N \partial_N A_M^{ij} + \partial_M \xi^N A_N^{ij} 
+ {\cal D}_M \theta^{ij} 
+ 2 i m \bar\epsilon^{[i} \psi_M^{j]}, \nonu
\delta A_M^{i'j'} 
\A = \A \xi^N \partial_N A_M^{i'j'} + \partial_M \xi^N A_N^{i'j'} 
+ {\cal D}_M \theta^{i'j'} 
- 2 i m \bar\epsilon^{[i'} \psi_M^{j']}. 
\label{localsym}
\ea
The transformation parameters $\xi^M(x)$, $\lambda^{AB}(x)$, 
$\theta^{ij}(x)$, $\theta^{i'j'}(x)$ and $\epsilon^i(x)$, 
$\epsilon^{i'}(x)$ represent general coordinate, local Lorentz, 
SO($p$) $\times$ SO($q$) gauge 
and local super transformations respectively. 
The parameters $\epsilon^i$, $\epsilon^{i'}$ are Majorana spinors 
and $\lambda^{AB} = -\lambda^{BA}$, $\theta^{ij} = -\theta^{ji}$, 
$\theta^{i'j'} = -\theta^{j'i'}$. 
The commutator algebra of these transformations closes for 
arbitrary $p$, $q$ modulo the field equations. 
\par
%
%%%%%%%  Section 3  %%%%%%%%%%%%%%%%%%%%%%%%%%%%%%%%%%%%%%%%
%
\newsection{Boundary behaviors of the fields}
It is convenient to partially fix the gauge for the local 
symmetries (\ref{localsym}). We represent the three-dimensional 
AdS space as a region $x^2 > 0$ in ${\bf R}^3$. 
The boundary of the AdS space corresponds to a plane $x^2 = 0$ and 
a point $x^2 = \infty$. We choose the gauge fixing condition as 
\ba
\A\A e_{M=2}{}^{A=2} = {1 \over 2mx^2}, \qquad 
e_{M=2}{}^a = 0, \qquad
e_\mu{}^{A=2} = 0, \nonu
\A\A \psi_2^i = 0, \qquad 
\psi_2^{i'} = 0, \qquad
A_2^{ij} = 0, \qquad
A_2^{i'j'} = 0, 
\label{gaugefix}
\ea
where $\mu, \nu, \cdots = 0,1$ and $a, b, \cdots = 0, 1$ are 
two-dimensional world indices and local Lorentz indices 
respectively. The metric in this gauge has a form 
\be
dx^M dx^N g_{MN} = {1 \over (2mx^2)^2} \left( 
dx^2 dx^2 + dx^\mu dx^\nu \hat g_{\mu\nu} \right). 
\label{metric}
\ee
The SO(2,2) invariant AdS metric corresponds to the case 
$\hat g_{\mu\nu} = \eta_{\mu\nu}$ but we consider the general 
$\hat g_{\mu\nu}$. We define $\hat e_\mu{}^a$ by 
$\hat g_{\mu\nu} = \hat e_\mu{}^a \hat e_\nu{}^b \eta_{ab}$. 
\par
Let us obtain asymptotic behaviors of the fields for 
$x^2 \rightarrow 0$. 
We assume that the dreibein $e_\mu{}^a$ behaves as $(x^2)^{-1}$ 
just as in the SO(2,2) invariant case. 
Asymptotic behaviors of other fields are determined by field 
equations. The field equations of the Rarita-Schwinger fields 
near $x^2 = 0$ are 
\be
\left( x^2 \partial_2 \pm {1 \over 2} \gamma_2 \right) \psi_\mu = 0, 
\ee
where $+$ is for $\psi_\mu = \psi_\mu^i$ and $-$ is for 
$\psi_\mu = \psi_\mu^{i'}$. The solutions behave as 
$\psi_{\mu\pm}^i \sim (x^2)^{\mp{1 \over 2}}$, 
$\psi_{\mu\pm}^{i'} \sim (x^2)^{\pm{1 \over 2}}$ for 
$x^2 \rightarrow 0$, where the suffices $\pm$ here denote eigenvalues 
of $\gamma^2$, i.e., chiralities in two-dimensional sense. 
The field equations which determine the boundary behavior of the 
gauge fields are 
\be
\partial_2 A_\mu = 0 
\ee
for both of $A_\mu = A_\mu^i, A_\mu^{i'}$. 
The solutions are independent of $x^2$. 
\par
According to the prescription in refs.\ \cite{GKP}, \cite{WITTEN} 
one has to impose boundary conditions on the fields. 
As for gravity we require that the zweibein $\hat e_\mu{}^a$ 
defined below eq.\ (\ref{metric}) approaches a given function 
$e_{0\mu}{}^a(x^0,x^1)$ at the boundary. 
Since the Rarita-Schwinger fields and the Chern-Simons gauge 
fields have field equations which are first order in 
derivatives, one should impose boundary conditions on only half 
of their components \cite{HS}, \cite{AF}. 
For the Rarita-Schwinger fields we impose boundary conditions 
on the components which become larger for $x^2 \rightarrow 0$, 
i.e., $\psi_{\mu+}^i$ and $\psi_{\mu-}^{i'}$. 
For the gauge fields all the components 
become independent of $x^2$ and one can choose either 
$A_- = e_-{}^\mu A_\mu$ or $A_+ = e_+{}^\mu A_\mu$. 
Here, the suffices $\pm$ denote the light-cone directions 
$e_\pm{}^\mu = {1 \over \sqrt{2}} ( e_0{}^\mu \pm e_1{}^\mu )$. 
We impose boundary conditions on $A_-^{ij}$ and $A_+^{i'j'}$. 
This choice is required by supersymmetry as we will see later. 
To summarize we impose boundary conditions on $e_\mu{}^a$, 
$\psi_{\mu+}^i$, $\psi_{\mu-}^{i'}$, $A_-^{ij}$ and $A_+^{i'j'}$. 
The boundary behaviors of these fields are 
\ba
\A\A e_\mu{}^a \rightarrow (2 m x^2)^{-1} e_{0\mu}{}^a, \nonu
\A\A \psi_{\mu +}^i \rightarrow (2 m x^2)^{-{1 \over 2}} 
\psi_{0\mu +}^i, \qquad
\psi_{\mu -}^{i'} \rightarrow (2 m x^2)^{-{1 \over 2}} 
\psi_{0\mu -}^{i'}, \nonu
\A\A A_-^{ij} \rightarrow 2 m x^2 A_{0-}^{ij}, \qquad
A_+^{i'j'} \rightarrow 2 m x^2 A_{0+}^{i'j'}, 
\label{bbehavior}
\ea
where the fields with the suffix $0$ are fixed functions 
on the boundary. Other components of the fields on the boundary 
are non-local functionals of the fields in eq.\ (\ref{bbehavior}), 
which are obtained by solving the field equations. 
We also introduce notations $\psi_{0\mu-}^i$, $\psi_{0\mu+}^{i'}$, 
$A_{0+}^i$, $A_{0-}^{i'}$ defined by 
$\psi_{\mu-}^i \rightarrow (2mx^2)^{1 \over 2} \psi_{0\mu-}^i$, etc. 
\par
%
%%%%%%%  Section 4  %%%%%%%%%%%%%%%%%%%%%%%%%%%%%%%%%%%%%%%%
%
\newsection{Local symmetries on the boundary}
Let us study how the fields on the boundary in eq.\ (\ref{bbehavior}) 
transform under the residual symmetry transformations after 
the gauge fixing. The residual symmetries, which preserve the gauge 
conditions (\ref{gaugefix}), are obtained by solving 
\ba
%\delta e_{M=2}{}^{A=2} 
\A\A \partial_2 \xi^2 - {1 \over x^2} \xi^2 = 0, \nonu
%
%\delta e_{M=2}{}^\mu 
\A\A \partial_2 \xi^\mu - \lambda^{a2} \hat e_a{}^\mu = 0, \nonu
%
%\delta e_\mu{}^{A=2} 
\A\A \lambda^2{}_a \hat e_\mu{}^a - \partial_\mu \xi^2 
- i m x^2 \left( \bar\epsilon^i \gamma^2 \psi_\mu^i 
+ \bar\epsilon^{i'} \gamma^2 \psi_\mu^{i'} \right) = 0, \nonu
%
%\delta \psi_{M=2}^i 
\A\A D_2 \epsilon^i + m \gamma_{M=2} \epsilon^i 
+ \partial_2 \xi^\mu \psi_\mu^i = 0, \nonu
%
%\delta \psi_{M=2}^{i'} 
\A\A D_2 \epsilon^{i'} - m \gamma_{M=2} \epsilon^{i'} 
+ \partial_2 \xi^\mu \psi_\mu^{i'} = 0, \nonu
%
%\delta A_{M=2}^{ij} 
\A\A \partial_2 \theta^{ij} 
+ \partial_2 \xi^\mu A_\mu^{ij} = 0, \nonu
%
%\delta A_{M=2}^{i'j'} 
\A\A \partial_2 \theta^{i'j'} 
+ \partial_2 \xi^\mu A_\mu^{i'j'} = 0. 
\label{residualeq}
\ea
These equations except the third one determine $x^2$-dependence 
of the transformation parameters. The third equation fixes 
$\lambda^{a2}$. 
The general solution of eq.\ (\ref{residualeq}) near the 
boundary $x^2 = 0$ is 
\ba
\xi^2 \A = \A - x^2 \Lambda_0(x^0, x^1), \qquad
\xi^\mu = \xi_0^\mu(x^0, x^1) + {\cal O}((x^2)^2), \nonu
\lambda^{ab} \A = \A \lambda_0^{ab}(x^0, x^1) + {\cal O}(x^2), \qquad
\lambda^{a2} = {\cal O}(x^2), \nonu
\epsilon_\pm^i \A = \A (2mx^2)^{\mp{1 \over 2}} \left[ 
\epsilon_{0\pm}^i(x^0, x^1) + {\cal O}(x^2) \right], \nonu
%
%\epsilon_-^i \A = \A (2mx^2)^{1 \over 2} \left[ 
%\epsilon_{0-}^i(x^0, x^1) + {\cal O}(x^2) \right], \nonu
%
\epsilon_\pm^{i'} \A = \A (2mx^2)^{\pm{1 \over 2}} \left[ 
\epsilon_{0\pm}^{i'}(x^0, x^1) + {\cal O}(x^2) \right], \nonu
%
%\epsilon_-^{i'} \A = \A (2mx^2)^{-{1 \over 2}} \left[ 
%\epsilon_{0-}^{i'}(x^0, x^1) + {\cal O}(x^2) \right], \nonu
%
\theta^{ij} \A = \A \theta_0^{ij}(x^0, x^1) + {\cal O}((x^2)^2), \qquad
\theta^{i'j'} = \theta_0^{i'j'}(x^0, x^1) + {\cal O}((x^2)^2), 
\label{residual}
\ea
where $\Lambda_0$, $\xi_0^\mu$, $\lambda_0^{ab}$, 
$\epsilon_{0\pm}^i$, $\epsilon_{0\pm}^{i'}$, $\theta_0^{ij}$ and 
$\theta_0^{i'j'}$ are arbitrary functions of $x^0$ and $x^1$. 
Order ${\cal O}(x^2)$ and ${\cal O}((x^2)^2)$ terms are non-local 
functionals of these functions and the fields. For instance, 
the order ${\cal O}((x^2)^2)$ term in $\theta^{ij}$ is given by 
$- \int_0^{x^2} d x^2 \partial_2 \xi^\mu A_\mu^{ij}$. 
Thus, the residual symmetry transformations of the fields 
in the bulk of the AdS space are non-local. 
\par
However, the transformations of the fields on the boundary 
in eq.\ (\ref{bbehavior}) can be local. Substituting 
eqs.\ (\ref{bbehavior}), (\ref{residual}) into eq.\ (\ref{localsym}) 
and taking the limit $x^2 \rightarrow 0$ we find the bosonic 
transformations of the fields on the boundary as 
\ba
\delta e_{0\mu}{}^a 
\A = \A \xi_0^\nu \partial_\nu e_{0\mu}{}^a 
+ \partial_\mu \xi_0^\nu e_{0\nu}{}^a + \Lambda_0 e_{0\mu}{}^a 
- \lambda_0^a{}_b e_{0\mu}{}^b, \nonu
\delta \psi_{0\mu+}^i 
\A = \A \xi_0^\nu \partial_\nu \psi_{0\mu+}^i 
+ \partial_\mu \xi_0^\nu \psi_{0\nu+}^i 
+ {1 \over 2} \Lambda_0 \psi_{0\mu+}^i 
- {1 \over 4} \lambda_0^{ab} \gamma_{ab} \psi_{0\mu+}^i 
- \theta_0^{ij} \psi_{0\mu+}^j, \nonu
\delta \psi_{0\mu-}^{i'} 
\A = \A \xi_0^\nu \partial_\nu \psi_{0\mu-}^{i'} 
+ \partial_\mu \xi_0^\nu \psi_{0\nu-}^{i'} 
+ {1 \over 2} \Lambda_0 \psi_{0\mu-}^{i'} 
- {1 \over 4} \lambda_0^{ab} \gamma_{ab} \psi_{0\mu-}^{i'} 
- \theta_0^{i'j'} \psi_{0\mu-}^{j'}, \nonu
\delta A_{0-}^{ij} 
\A = \A \xi_0^\nu \partial_\nu A_{0-}^{ij} 
- \Lambda_0 A_{0-}^{ij} - \lambda_-{}^- A_{0-}^{ij} 
+ {\cal D}_{0-} \theta_0^{ij}, \nonu
\delta A_{0+}^{i'j'} 
\A = \A \xi_0^\nu \partial_\nu A_{0+}^{i'j'} 
- \Lambda_0 A_{0+}^{i'j'} - \lambda_+{}^+ A_{0+}^{i'j'} 
+ {\cal D}_{0+} \theta_0^{i'j'}. 
\label{bbtrans}
\ea
We see that the transformations with the parameters $\xi_0^\mu$, 
$\Lambda_0$, $\lambda_0^{ab}$ and $\theta_0^{ij}$, $\theta_0^{i'j'}$ 
represent general coordinate, Weyl, local Lorentz and 
SO($p$) $\times$ SO($q$) gauge transformations in two dimensions 
respectively. 
In particular, the general coordinate transformation in the 
direction $M=2$ became two-dimensional Weyl transformation. 
Weights of the Weyl transformation are determined by the powers 
of $x^2$ appearing in the boundary behaviors of the fields 
(\ref{bbehavior}). 
\par
%
%%%%%%%%%%%%%%%%%%%%%%%%%%%%%%%%%%%%%%%%%%%%%%%%%%%%%%%%%%%%
%
On the other hand, in the limit $x^2 \rightarrow 0$ the 
fermionic transformations of the fields on the boundary 
in eq.\ (\ref{bbehavior}) become 
\ba
\delta e_{0\mu}{}^a 
\A = \A {1 \over 2} i \left( \bar\epsilon_{0+}^i \gamma^a 
\psi_{0\mu+}^i 
+ \bar\epsilon_{0-}^{i'} \gamma^a \psi_{0\mu-}^{i'} \right), \nonu
\delta \psi_{0\mu+}^i 
\A = \A D_{0\mu} \epsilon_{0+}^i 
+ A_{0\mu}^{ij} \epsilon_{0+}^j 
+ 2m \gamma_{0\mu} \epsilon_{0-}^i, \nonu
\delta \psi_{0\mu-}^{i'} 
\A = \A D_{0\mu} \epsilon_{0-}^{i'} 
+ A_{0\mu}^{i'j'} \epsilon_{0-}^{j'} 
- 2m \gamma_{0\mu} \epsilon_{0+}^{i'}, \nonu
\delta A_{0-}^{ij} 
\A = \A 2 i m e_{0-}{}^\mu \left( \bar\epsilon_{0-}^{[i} 
\psi_{0\mu+}^{j]} 
+ \bar\epsilon_{0+}^{[i} \psi_{0\mu-}^{j]} \right) 
+ \delta e_{0-}{}^\mu A_{0\mu}^{ij}, \nonu
\delta A_{0+}^{i'j'} 
\A = \A - 2 i m e_{0+}{}^\mu \left( \bar\epsilon_{0-}^{[i'} 
\psi_{0\mu+}^{j']} 
+ \bar\epsilon_{0+}^{[i'} \psi_{0\mu-}^{j']} \right) 
+ \delta e_{0+}{}^\mu A_{0\mu}^{i'j'}. 
\label{bftrans}
\ea
The transformation of $e_{0\mu}{}^a$ is that of the two-dimensional 
($p, q$) supergravities, i.e., supergravities with 
$p$ supersymmetries of positive chirality and $q$ of 
negative chirality (See, e.g., ref.\ \cite{SS}.). 
However, the transformations of other fields have different forms 
from those of the two-dimensional supergravities. 
Furthermore, they contain $\psi_{0\mu-}^i$, $\psi_{0\mu+}^{i'}$, 
$A_+^i$, $A_-^{i'}$, which are non-local functionals of the 
fields in eq.\ (\ref{bbehavior}). 
We shall try to rewrite these transformations in a local form 
by using field equations. 
\par
Using an identity 
\be
\eta_{ab} = {1 \over 2} ( \eta_{ab} + \epsilon_{ab} \gamma^2 ) 
+ {1 \over 2} \gamma_a \gamma_b 
\ee
in the second terms, the transformations of the Rarita-Schwinger 
fields in eq.\ (\ref{bftrans}) can be rewritten as 
\ba
\delta \psi_{0\mu+}^i \A = \A D_{0\mu} \epsilon_{0+}^i 
+ e_{0\mu}{}^- A_{0-}^{ij} \epsilon_{0+}^j 
+ \gamma_{0\mu} \eta_{0-}^i, \nonu
\delta \psi_{0\mu-}^{i'} \A = \A D_{0\mu} \epsilon_{0-}^{i'} 
+ e_{0\mu}{}^+ A_{0+}^{i'j'} \epsilon_{0-}^{j'} 
+ \gamma_{0\mu} \eta_{0+}^{i'}, 
\label{RStrans}
\ea
where 
\ba
\eta_{0-}^i \A = \A 2m \epsilon_{0-}^i 
+ {1 \over 2} A_{0+}^{ij} \gamma^+ \epsilon_{0+}^j, \nonu
\eta_{0+}^{i'} \A = \A -2m \epsilon_{0+}^{i'} 
+ {1 \over 2} A_{0-}^{i'j'} \gamma^- \epsilon_{0-}^{j'}. 
\ea
If we regard $\eta_{0-}^i$, $\eta_{0+}^{i'}$ as independent 
transformation parameters, eq.\ (\ref{RStrans}) do not contain 
non-local functionals anymore. 
\par
As for the gauge fields we use field equations of the 
Rarita-Schwinger fields to eliminate $\psi_{0\mu-}^i$ and 
$\psi_{0\mu+}^{i'}$. First, the first term of the transformation 
of $A_{0-}^{ij}$ in eq.\ (\ref{bftrans}) can be rewritten as 
\be
i m \left( \bar\epsilon_{0+}^{[i} \gamma_- 
\gamma_0^\nu \psi_{0\nu-}^{j]} 
+ \bar\epsilon_{0-}^{[i} \gamma_0^\nu 
\gamma_- \psi_{0\nu+}^{j]} \right). 
\label{firstterm}
\ee
{}From the Rarita-Schwinger field equations we obtain 
\be
m \gamma_{0\mu} \gamma_0^\nu \psi_{0\nu-}^i 
= {1 \over 2} \gamma_0^\nu \psi_{0\mu\nu+}^i 
- {1 \over 4} \gamma_{0\mu} \gamma_0^\nu \slA_0^{ij} \psi_{0\nu+}^j, 
\label{gammatrace}
\ee
where 
\ba
\psi_{0\mu\nu+}^i \A = \A D_{0\mu} \psi_{0\nu+}^i 
+ e_{0\mu}{}^- A_{0-}^{ij} \psi_{0\nu+}^j 
- (\mu \leftrightarrow \nu), \nonu 
\psi_{0\mu\nu-}^{i'} \A = \A D_{0\mu} \psi_{0\nu-}^{i'} 
+ e_{0\mu}{}^+ A_{0+}^{i'j'} \psi_{0\nu-}^{j'} 
- (\mu \leftrightarrow \nu). 
\ea
($\psi_{0\mu\nu-}^{i'}$ is for later use.) 
Substituting eq.\ (\ref{gammatrace}) into eq.\ (\ref{firstterm}) 
we obtain an expression for $\delta A_{0-}^{ij}$ independent of 
$\psi_{0\mu-}^i$. Similarly we obtain an expression for 
$\delta A_{0\mu+}^{i'j'}$ independent of $\psi_{0\mu+}^{i'}$. 
Thus the transformations of the gauge fields become 
\ba
\delta A_{0-}^{ij} 
\A = \A {1 \over 2} i \bar\epsilon_{0+}^{[i} \gamma_0^\nu 
\psi_{0\mu\nu+}^{j]} e_{0-}{}^\mu 
+ {1 \over 2} i \bar\eta_{0-}^{[i} \gamma_0^\mu 
\gamma_- \psi_{0\mu+}^{j]} 
- {1 \over 2} i \bar\epsilon_{0-}^{k'} \gamma^- \psi_{0\mu-}^{k'} 
e_{0-}{}^\mu A_{0-}^{ij} \nonu
\A \A - {3 \over 4} i A_{0+}^{[ij} \bar\epsilon_{0+}^{k]} 
\gamma^+ \psi_{0\mu+}^k e_{0-}{}^\mu 
- {3 \over 4} i \bar\epsilon_{0+}^k 
\gamma^+ \psi_{0\mu+}^{[i} A_{0+}^{jk]} e_{0-}{}^\mu, \nonu
\delta A_{0+}^{i'j'} 
\A = \A {1 \over 2} i \bar\epsilon_{0-}^{[i'} \gamma_0^\nu 
\psi_{0\mu\nu-}^{j']} e_{0+}{}^\mu 
+ {1 \over 2} i \bar\eta_{0+}^{[i'} \gamma_0^\mu 
\gamma_+ \psi_{0\mu-}^{j']} 
- {1 \over 2} i \bar\epsilon_{0+}^k \gamma^+ \psi_{0\mu+}^k 
e_{0+}{}^\mu A_{0+}^{i'j'} \nonu
\A \A - {3 \over 4} i A_{0-}^{[i'j'} \bar\epsilon_{0-}^{k']} 
\gamma^- \psi_{0\mu-}^{k'} e_{0+}{}^\mu 
- {3 \over 4} i \bar\epsilon_{0-}^{k'} 
\gamma^- \psi_{0\mu-}^{[i'} A_{0-}^{j'k']} e_{0+}{}^\mu. 
\label{Atrans}
\ea
The last two terms in these transformations still contain the fields 
$A_{0+}^{ij}$ or $A_{0-}^{i'j'}$, which are non-local functionals 
of the boundary fields. These terms vanish for $p, q \leq 2$ since 
three indices $i, j, k$ or $i', j', k'$ are antisymmetrized. 
Therefore, we have a local form of fermionic transformations 
only for $p$, $q \leq 2$. 
\par
%
%%%%%%%  Section 5  %%%%%%%%%%%%%%%%%%%%%%%%%%%%%%%%%%%%%%%%
%
\newsection{Comparison with two-dimensional conformal supergravities}
Let us compare the above fermionic transformations of the boundary 
fields (\ref{bftrans}), (\ref{RStrans}), (\ref{Atrans}) obtained from 
the AdS/CFT correspondence with those in the two-dimensional 
($p, q$) conformal supergravities for $p$, $q \leq 2$. 
We begin with the case $p=q=2$. 
The two-dimensional ($2,2$) conformal supergravity \cite{BS} 
contains a zweibein $\tilde e_\mu{}^a$, Majorana Rarita-Schwinger 
fields $\tilde\psi_\mu^i$ ($i = 1, 2$) and a real vector field 
$\tilde A_\mu^{ij}$. Their fermionic transformations are 
\ba
\delta \tilde e_\mu{}^a 
\A = \A {1 \over 2} \, i \, \bar{\tilde\epsilon}^i 
\gamma^a \tilde\psi_\mu^i, \qquad
\delta \tilde\psi_\mu^i 
= \tilde D_\mu \tilde\epsilon^i 
+ \tilde A_\mu^{ij} \tilde\epsilon^j 
+ \tilde\gamma_\mu \tilde\eta^i, \nonu
\delta \tilde A_\mu^{ij} 
\A = \A {1 \over 2} \, i \, \bar{\tilde\epsilon}^{[i} \tilde\gamma_\mu 
\tilde\gamma^{\rho\sigma} \left( \tilde D_\rho \tilde\psi_\sigma^{j]} 
+ \tilde A_\rho^{j]k} \tilde\psi_\sigma^k \right) 
+ {1 \over 2} \, i \, \bar{\tilde\eta}^{[i} 
\tilde\gamma^\nu \tilde\gamma_\mu \tilde\psi_\nu^{j]}, 
\label{22sugra}
\ea
where the transformation parameters $\tilde\epsilon^i$ and 
$\tilde\eta^i$ are Majorana spinors and represent the 
supertransformation and the super Weyl transformation respectively. 
By identifying the fields (\ref{bbehavior}) with these fields as 
\ba
\tilde A_\mu^{ij} 
\A = \A e_{0\mu}{}^- A_{0-}^{ij} 
+ e_{0\mu}{}^+ A_{0+}^{i'j'}, \qquad
\tilde \psi_\mu^i 
= \psi_{0\mu+}^i + \psi_{0\mu-}^{i'}, \nonu
\tilde\epsilon^i 
\A = \A \epsilon_{0+}^i + \epsilon_{0-}^{i'}, \qquad
\tilde\eta^i 
= \eta_{0-}^i + \eta_{0+}^{i'} 
- {1 \over 2} \slA_0^{i'j'} \epsilon_{0+}^j 
- {1 \over 2} \slA_0^{ij} \epsilon_{0-}^{j'}, 
\ea
where $i' = i$, $j'= j$, the transformations obtained from the 
AdS/CFT correspondence can be shown to reproduce the fermionic 
transformations in eq.\ (\ref{22sugra}). 
\par
The fermionic transformations of the ($2,1$) conformal supergravity 
can be obtained from those of the ($2,2$) theory (\ref{22sugra}) 
by a truncation 
\be
\tilde\psi_{\mu-}^2 = 0, \qquad
\tilde A_+^{12} = 0, \qquad
\tilde\epsilon_-^2 = 0, \qquad
\tilde\eta_+^2 = {1 \over 2} \tilde\slA^{12} \tilde\epsilon_-^1. 
\label{21tranc}
\ee
The transformations of the remaining fields $\tilde e_\mu{}^a$, 
$\tilde\psi_{\mu+}^i$, $\tilde\psi_{\mu-}^2$, $\tilde A_-^{12}$ 
are exactly the same as those obtained from the AdS/CFT 
correspondence by an obvious identification of the fields. 
The ($1,1$) theory \cite{DZ} contains $\tilde e_\mu{}^a$, 
$\tilde\psi_\mu^1$, whose fermionic transformations are obtained 
from the ($2,2$) theory by a truncation $\tilde\psi_\mu^2 = 0$, 
$\tilde A_\mu^{12} = 0$, $\tilde\epsilon^2 = 0$, $\tilde\eta^2 = 0$. 
On the other hand, the ($2,0$) theory \cite{BSN} 
is obtained from the ($2,2$) theory by a truncation 
$\tilde\psi_{\mu-}^i = 0$, $\tilde A_+^{12} = 0$, 
$\tilde\epsilon_-^i = 0$, $\tilde\eta_+^i = 0$. 
The ($1,0$) theory \cite{GHMR} is obtained from 
the ($2,0$) theory by further truncation 
$\tilde\psi_{\mu+}^2 = 0$, $\tilde A_-^{12} = 0$, 
$\tilde\epsilon_+^2 = 0$, $\tilde\eta_-^2 = 0$. 
The fermionic transformations of these theories coincide with 
those obtained from the AdS/CFT correspondence. 
\par
Thus, for all $p, q \leq 2$ the fermionic transformations of the 
boundary fields are locally realized and are exactly the same 
as the super and the super Weyl transformations of two-dimensional 
($p, q$) conformal supergravities. 
For $p > 2$ or $q > 2$ the fermionic transformations of 
the gauge fields are non-local and a relation to two-dimensional 
conformal supergravities is not clear. 
We note here that the construction of the two-dimensional ($p, p$) 
conformal supergravities based on the super Lie algebra 
OSp(2, $p$) $\oplus$ OSp(2, $p$) in ref.\ \cite{BSN} also 
failed for $p > 2$. It would be interesting to see a relation 
between this construction and the AdS/CFT correspondence. 

\vspace{10mm}

%\newpage
%
%%%%%  References  %%%%%%%%%%%%%%%%%%%%%%%%%%%%%%%%%%%%%%%
%
\newcommand{\NP}[1]{{\it Nucl.\ Phys.\ }{\bf #1}}
\newcommand{\PL}[1]{{\it Phys.\ Lett.\ }{\bf #1}}
\newcommand{\CMP}[1]{{\it Commun.\ Math.\ Phys.\ }{\bf #1}}
\newcommand{\MPL}[1]{{\it Mod.\ Phys.\ Lett.\ }{\bf #1}}
\newcommand{\IJMP}[1]{{\it Int.\ J. Mod.\ Phys.\ }{\bf #1}}
\newcommand{\PR}[1]{{\it Phys.\ Rev.\ }{\bf #1}}
\newcommand{\PRL}[1]{{\it Phys.\ Rev.\ Lett.\ }{\bf #1}}
\newcommand{\PTP}[1]{{\it Prog.\ Theor.\ Phys.\ }{\bf #1}}
\newcommand{\PTPS}[1]{{\it Prog.\ Theor.\ Phys.\ Suppl.\ }{\bf #1}}
\newcommand{\AP}[1]{{\it Ann.\ Phys.\ }{\bf #1}}

\begin{thebibliography}{100}
%
\bibitem{MAL} J. Maldacena, 
        The large $N$ limit of superconformal field theories 
        and supergravity, hep-th/9711200. 
\bibitem{GKP} S.S. Gubser, I.R. Klebanov and A.M. Polyakov, 
        \PL{B428} (1998) 105, hep-th/9802109. 
\bibitem{WITTEN} E. Witten, 
        Anti de Sitter space and holography, hep-th/9802150. 
%
\bibitem{HS} M. Henningson and K. Sfetsos, \PL{B431} (1998) 63, 
        hep-th/9803251. 
\bibitem{AF} G.E. Arutyunov and S.A. Frolov, On the origin of 
        supergravity boundary terms in the AdS/CFT correspondence, 
        hep-th/9806216. 
%
\bibitem{FFZ} S. Ferrara, C. Fr\o nsdal and A. Zaffaroni, 
        On $N=8$ supergravity on ${\rm AdS}_5$ and $N=4$ 
        superconformal Yang-Mills theory, hep-th/9802203. 
\bibitem{LT} H. Liu and A.A. Tseytlin, 
        $D=4$ super Yang-Mills, $D=5$ gauged supergravity 
        and $D=4$ conformal supergravity, hep-th/9804083. 
%
\bibitem{AT} A. Ach\'ucarro and P.K. Townsend, \PL{B180} (1986) 89. 
%
\bibitem{BH} J.D. Brown and M. Henneaux, \CMP{104} (1986) 207. 
\bibitem{BBCHO} M. Ba\~nados, K. Bautier, O. Coussaert, M. Henneaux 
        and M. Ortiz, \PR{D58} (1998) 085020, hep-th/9805165. 
\bibitem{BOER} J. de Boer, Six-dimensional supergravity on 
        ${\rm S}^3 \times {\rm AdS}_3$ and 2d conformal field theory, 
        hep-th/9806104. 
\bibitem{GKS} A. Giveon, D. Kutasov and N. Seiberg, Comments on 
        string theory on ${\rm AdS}_3$, hep-th/9806194. 
%
\bibitem{SS} A. Salam and E. Sezgin, 
        {\it Supergravities in Diverse Dimensions} 
        (North-Holland/World Scientific, 1989). 
\bibitem{BS} L. Brink and J.H. Schwarz, \NP{B121} (1977) 285. 
\bibitem{DZ} S. Deser and B. Zumino, \PL{B65} (1976) 369; 
        L. Brink, P. Di Vecchia and P. Howe, \PL{B65} (1976) 471. 
\bibitem{BSN} E. Bergshoeff, E. Sezgin and H. Nishino, 
        \PL{B166} (1986) 141. 
\bibitem{GHMR} D.J. Gross, J.A. Harvey, E. Martinec and R. Rhom, 
        \NP{B256} (1985) 253. 
%
\end{thebibliography}
\end{document}